\documentclass[12pt,preprint]{aastex}
\usepackage{natbib}
\usepackage{subfigure}
\usepackage{lscape}

\newcommand{\captionfonts}{\small}

\makeatletter  % Allow the use of @ in command names
\long\def\@makecaption#1#2{%
  \vskip\abovecaptionskip
  \sbox\@tempboxa{{\captionfonts #1: #2}}%
  \ifdim \wd\@tempboxa >\hsize
    {\captionfonts #1: #2\par}
  \else
    \hbox to\hsize{\hfil\box\@tempboxa\hfil}%
    \fi
  \vskip\belowcaptionskip}
\makeatother   % Cancel the effect of \makeatletter

\defcitealias{kgb}{KGB}

\begin{document}

\title{The Relationship Between Molecular Gas and Star Formation in
  Low-Mass E/S0 Galaxies}

\author{Lisa H. Wei\altaffilmark{1}, Stuart N. Vogel\altaffilmark{1},
  Sheila J. Kannappan\altaffilmark{2}, Andrew
  J. Baker\altaffilmark{3}, David V. Stark\altaffilmark{2}, Seppo
  Laine\altaffilmark{4}} \altaffiltext{1}{Department of Astronomy,
  University of Maryland, College Park, MD 20742-2421}
  \altaffiltext{2}{Department of Physics and Astronomy, University of
  North Carolina, Phillips Hall CB 3255, Chapel Hill, NC 27599-3255}
  \altaffiltext{3}{Department of Physics and Astronomy, Rutgers, the
  State University of New Jersey, 136 Frelinghuysen Road, Piscataway,
  NJ 08854-8019} \altaffiltext{4}{\textit{Spitzer} Science Center,
  California Institute of Technology, MS 220-6, Pasadena, CA 91125}

\begin{abstract}

  We consider the relationship between molecular-gas and
  star-formation surface densities in 19 morphologically defined E/S0s
  with stellar mass $\la4\times10^{10}\,M_\odot$, paying particular
  attention to those found on the blue sequence in color vs. stellar
  mass parameter space, where spiral galaxies typically reside. While
  some blue-sequence E/S0s must be young major-merger remnants, many
  low-mass blue-sequence E/S0s appear much less disturbed, and may be
  experiencing the milder starbursts associated with inner-disk
  building as spirals (re)grow. For a sample of eight E/S0s (four
  blue-, two mid-, and two red-sequence) whose CARMA CO(1--0), {\it
    Spitzer} MIPS $24\,{\rm \mu m}$, and {\it GALEX} FUV emission
  distributions are spatially resolved on a 750\,pc scale, we find
  roughly linear relationships between molecular-gas and
  star-formation surface densities within all galaxies, with power law
  indices $N =\,$~0.6--1.9 (median 1.2). Adding 11 more blue-sequence
  E/S0s whose CO(1--0) emission is not as well resolved, we find that
  most of our E/S0s have global 1--8 kpc aperture-averaged
  molecular-gas surface densities overlapping the range spanned by the
  disks and centers of spiral galaxies. While many of our E/S0s fall
  on the same Schmidt-Kennicutt relation as local spirals, $\sim$80\%
  (predominantly on the blue sequence) are offset towards apparently
  higher molecular-gas star formation efficiency (i.e., shorter
  molecular gas depletion time). Possible interpretations of the
  elevated efficiencies include bursty star formation similar to that
  in local dwarf galaxies, H$_2$ depletion in advanced starbursts, or
  simply a failure of the CO(1--0) emission to trace all of the
  molecular gas.

\end{abstract}

\keywords{galaxies: elliptical and lenticular, cD --- galaxies:
  evolution --- galaxies: star formation}

\section{Introduction}\label{section.intro}

Much progress has been made in recent decades in understanding the
relationship between star formation rate (SFR) surface density
($\Sigma_{\rm SFR}$) and gas surface density ($\Sigma_{\rm gas}$) in
nearby star-forming spiral galaxies, and this work has been used to
infer the physical basis of the star formation law (e.g.,
\citealt{krumholz09,murray09}). Studies relating the two observables
with a power law of the form

\begin{equation}
\Sigma_{\rm SFR}\,\,=\,\,a\Sigma_{\rm gas}^N,
\end{equation}

\noindent typically find power-law indices \textit{N} ranging from 1
to 3 (e.g., \citealt{kennicutt98,wong02,kennicutt07}; see
\citealt{bigiel08} for a review of previous work). In a spatially
resolved study of HI and H$_2$ in star-forming spirals,
\citet{bigiel08} find a linear ($N\,\sim\,$1) relation between
$\Sigma_{\rm SFR}$ and molecular-gas surface density ($\Sigma_{\rm
H_{2}}$), but little to no correlation between $\Sigma_{\rm SFR}$ and
HI surface density.

In contrast, less is known about the connection between $\Sigma_{\rm
  gas}$ and star formation in early-type galaxies. The good
  correlation between the morphologies of molecular gas, 24\,$\mu$m
  emission, and radio continuum in local E/S0s hints that there is a
  relationship between $\Sigma_{\rm H_{2}}$ and $\Sigma_{\rm SFR}$
  \citep{young09}. In a single-dish survey of CO emission in SAURON
  E/S0s, \citet{combes07} find that their galaxies follow the
  $N\,=\,1.4$ disk-averaged power law, characteristic of spirals
  \citep{kennicutt98}. Using multiple star-formation tracers,
  \citet{crocker10} find a similar result for a sample of 12 E/S0s,
  although possibly at lower total-gas star formation efficiencies
  (TSFE $\equiv$ SFR/$M_{\rm HI+H_2+He}$). \citet{shapiro10} update
  the \citeauthor{combes07} results with spatially resolved maps,
  localizing both CO emission and star formation in the central
  regions of the galaxies and finding $\Sigma_{\rm SFR}$, $\Sigma_{\rm
  H_{2}}$, and molecular-gas star formation efficiency (MSFE $\equiv$
  SFR/$M_{\rm H_2+He}$) values similar to those of spirals. However,
  spatially resolved studies of the H$_2$-star formation relation at
  sub-kpc resolution similar to the analyses of \citet{kennicutt07}
  and \citet{bigiel08} have yet to be done for E/S0s.

Recent work has identified a local population of star-forming E/S0s
that reside alongside spirals on the blue sequence in color
vs. stellar mass space \citep*[hereafter KGB]{kgb}. The fraction of
E/S0s on the blue sequence increases with decreasing mass, from $\ga
5\%$ at stellar mass $M_*\,\sim 3\times 10^{10}\,M_{\odot}$, up to
$\ga 20-30\%$ for $M_*\,\la 5\times 10^{9}\,M_{\odot}$ (2\% and 5\% ,
respectively, of all galaxies in these mass ranges;
\citetalias{kgb}). High-mass blue-sequence E/S0s are often young
major-merger remnants that will fade to the red sequence (see also
\citealt{schawinski09}). Low-mass blue-sequence E/S0s, in contrast,
appear more settled, occupying low-density field environments where
gas accretion is likely \citepalias{kgb}. This population may reflect
the transformation of ``red and dead'' E/S0s into spirals via inner-
and outer-disk regrowth \citepalias{kgb}. \citetalias{kgb} argue that
many blue-sequence E/S0s occupy a ``sweet spot'' in $M_*$ and stellar
concentration index, characterized by abundant gas and bursty,
efficient (when time-averaged over multiple bursts) star formation,
which may enable efficient disk building (see also \S
\ref{section.discussion}). \citet{wei10} confirm that blue-sequence
E/S0s have fractionally large atomic gas reservoirs, comparable to
those of spirals (0.1--1.0, relative to $M_*$). The ongoing star
formation and large gas reservoirs of these galaxies make them ideal
for probing the spatially resolved relationship between $\Sigma_{\rm
H_{2}}$ and $\Sigma_{\rm SFR}$ in E/S0s, offering unique insight into
whether/how some E/S0s may be actively evolving via bursty, efficient
star formation.

\section{Sample \& Data}\label{section.sample}

\begin{figure}
\includegraphics[scale=1]{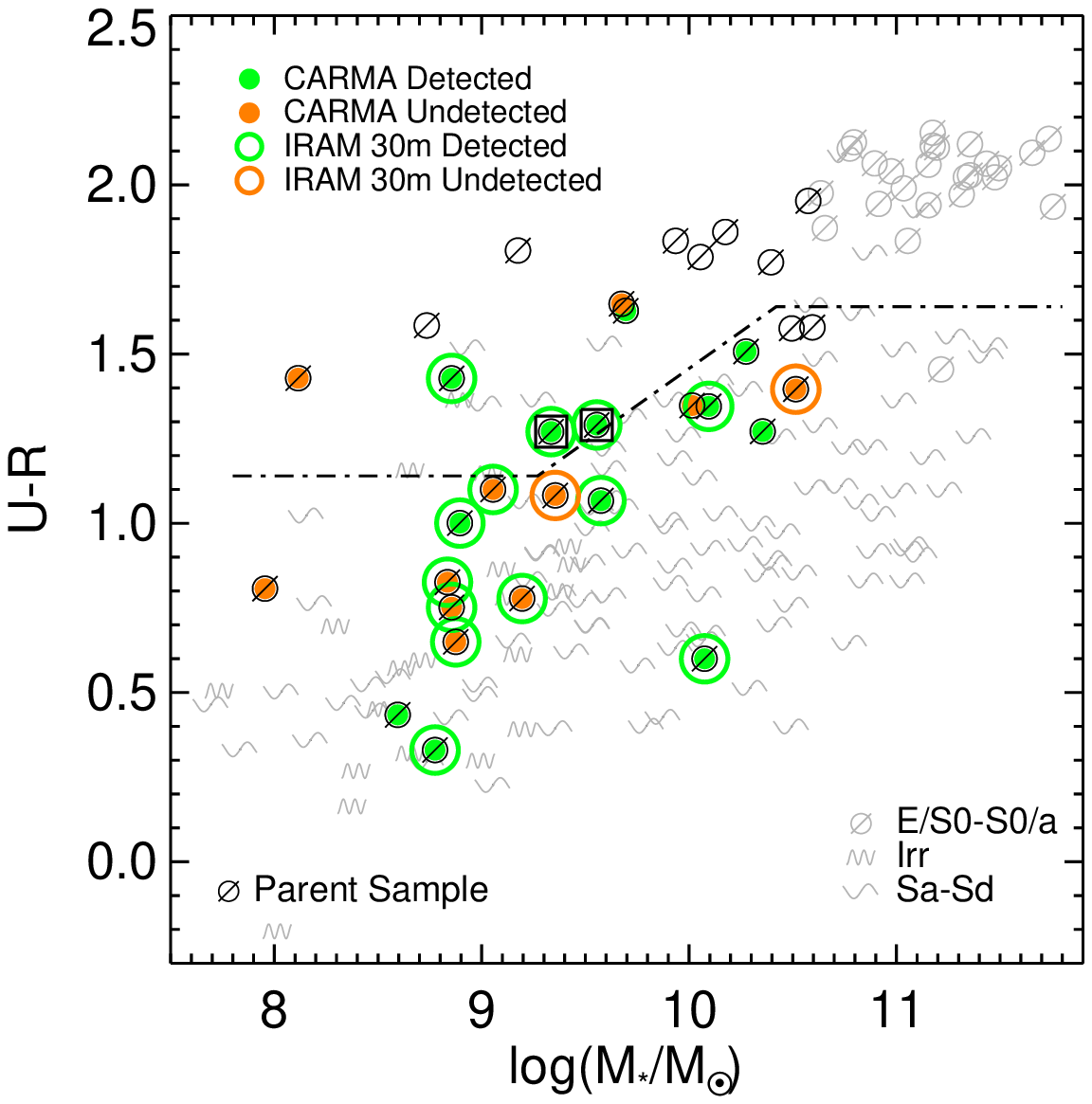}
\caption{$U-R$ color vs. stellar mass for galaxies in the Nearby Field
  Galaxy Survey \citep{jansen00b}, plus five additional E/S0s from the
  literature. Symbols denote morphological types \citepalias{kgb}. The
  red sequence, i.e. the main locus of traditional red E/S0s, lies
  above the dashed line (with two borderline ``mid-sequence'' E/S0s
  boxed, see \citetalias{kgb}), while the blue sequence (typically
  populated by spirals) lies below. Dark symbols denote the 32
  galaxies in the parent sample; the rest of the NFGS is shown in
  light grey. \label{fig.sample}}
\end{figure}

Our parent sample of 32 galaxies consists of all E/S0s (14 blue-, 2
mid-, and 11 red-sequence) with ${M_* \le 4 \times
10^{10}\,{M_{\odot}}}$ from the Nearby Field Galaxy Survey (NFGS,
\citealt{jansen00b}), and an additional five blue-sequence E/S0s in
the same mass range (see \citetalias{kgb}). The sample was defined for
\textit{Spitzer} and \textit{GALEX} programs GO-30406 and GI3-046012
(PI Kannappan), so all galaxies have new or archival FUV and
24\,$\mu$m data for SFR estimation. Of the parent sample, we observed
23 E/S0s in CO(1--0) with the Combined Array for Research in
Millimeter-Wave Astronomy (CARMA), detecting 12 of 23. In this Letter,
we focus on the 12 E/S0s with CARMA detections and 7 additional E/S0s
with IRAM 30\,m observations (five detections, two limits) in
CO(1--0). These 19 E/S0s sample a large range
(Table~\ref{table.global}) in color and stellar mass
(Figure~\ref{fig.sample}), H$_2$/HI mass ratio (0.006--3.2), and total
gas-to-stellar mass ratio (0.07--3.4).

% For galaxies showing fairly regular CO rotation,
%we fit velocity maps with model rotation curves and created masks from
%adaptive windows around the model velocity fields to make
%integrated-velocity maps. The four galaxies (NGC~3773, NGC~5338,
%UGC~6003, UGC~6570) without regular rotation in CO were masked in
%velocity based on their HI velocity profiles.

The CARMA maps have beam sizes of 2$\arcsec$--4$\arcsec$, with
velocity coverage of 300--450 $\rm km\,s^{-1}$ and resolution of
$\sim$2.5 $\rm km\,s^{-1}$. We reduced the data with the MIRIAD
package \citep{sault95}, using natural weighting. A more detailed
description of the CARMA data reduction will be provided in L. Wei et
al. 2010, in preparation. IRAM 30\,m CO(1--0) observations are from
S. Kannappan et al. 2010, in preparation, and D. Stark et al. 2010, in
preparation, with additional literature data as noted in Table
\ref{table.global}. We consider central pointings only, so the
23$\arcsec$ IRAM beam probes inner disks (relative to
$\sim$0$\farcm$6--1$\farcm$9 optical diameters). Comparison between
CARMA and IRAM 30\,m fluxes suggests that little flux is resolved out
by the interferometric observations.

We use the pipeline 24\,$\mu$m mosaics from the \textit{Spitzer}
archive, and the background-subtracted pipeline-processed FUV data
from the \textit{GALEX} archive. The CARMA and \textit{GALEX} data
sets were convolved with a kernel that reproduce the MIPS 24\,$\mu$m
PSF ($\rm FWHM$ $\sim 6\arcsec$; \citealt{gordon08}) exactly in the
CARMA/\textit{GALEX} images, including the Airy ring, which is
$<\,1\%$ of the $I_{24,\rm peak}$. For our pixel-to-pixel analysis in
\S \ref{section.pixels}, we further convolve our data to a resolution
of 750\,pc (6.1$\arcsec$--15.3$\arcsec$) for comparison with
\citet{bigiel08}. All maps are at least Nyquist-sampled.

We estimate ${\rm \Sigma_{H_2}}$ and ${\rm \Sigma_{SFR}}$ following
\citet{leroy08}:

\begin{equation}
\label{eq.h2}
{\frac {\Sigma_{\rm H_2}}{M_\odot\,{\rm pc}^{-2}}}  =  4.4\,{\rm 
cos}\,i\,{\frac {I_{\rm CO(1-0)}}{\rm K\,km\,s^{-1}}} 
\end{equation}

\begin{equation}
\label{eq.sfr}
{\frac {\Sigma_{\rm SFR}}{M_\odot\,{\rm kpc}^{-2}\,{\rm yr^{-1}}}}  = 
{\rm cos}\,i\,{\frac {8.1 \times 10^{-2}\,I_{\rm FUV} +
3.2 \times 10^{-3}\,I_{24}}{\rm MJy\,sr^{-1}}},
\end{equation}

\noindent assuming a CO-to-H$_{2}$ conversion factor ($X_{\rm CO}$) of
${\rm 2 \times 10^{20}\, cm^{-2}\,\,(K\,\, km\,\, s^{-1})^{-1}}$. Our
estimates of $\Sigma_{\rm H_{2}}$ do not include helium; however,
helium is included in estimates of molecular gas depletion
time. Equation \ref{eq.sfr} uses the broken power law initial mass
function given by \citet{kroupa01}. Our procedures for calculating
${\rm \Sigma_{H_2}}$ and ${\rm \Sigma_{SFR}}$ are identical to those
of \citet{bigiel08} after their conversion from CO(2--1) to CO(1--0)
($I_{\rm CO(2-1)}$/$I_{\rm CO(1-0)} = 0.8$), ensuring a fair
comparison between the two data sets.

AGN contamination is not an issue for the 750\,pc resolution analysis,
as any AGN contribution in the infrared would be contained within the
central resolution element. NGC~4117 and IC~1141 are known AGN hosts,
and optical line diagnostics from \citet{kewley06} identify NGC~5173
as a candidate host. However, IRAC color-color diagnostics
\citep{sajina05} suggest that (possible) AGN contributions to the
integrated infrared emission in our galaxies are relatively weak, as
the removal of the central resolution element does not significantly
affect their positions in the infrared color-color diagram. Following
\citet{temi07}, we estimate possible 24\,$\mu$m emission contamination
from passively evolving stellar populations to be $<$\,8\% for our
E/S0s.

\section{The Resolved Star Formation Relation at 750\,pc
  Resolution}\label{section.pixels}

\begin{figure}
\vspace{-.5cm}
\hspace{-.75cm}
\includegraphics[scale=.85]{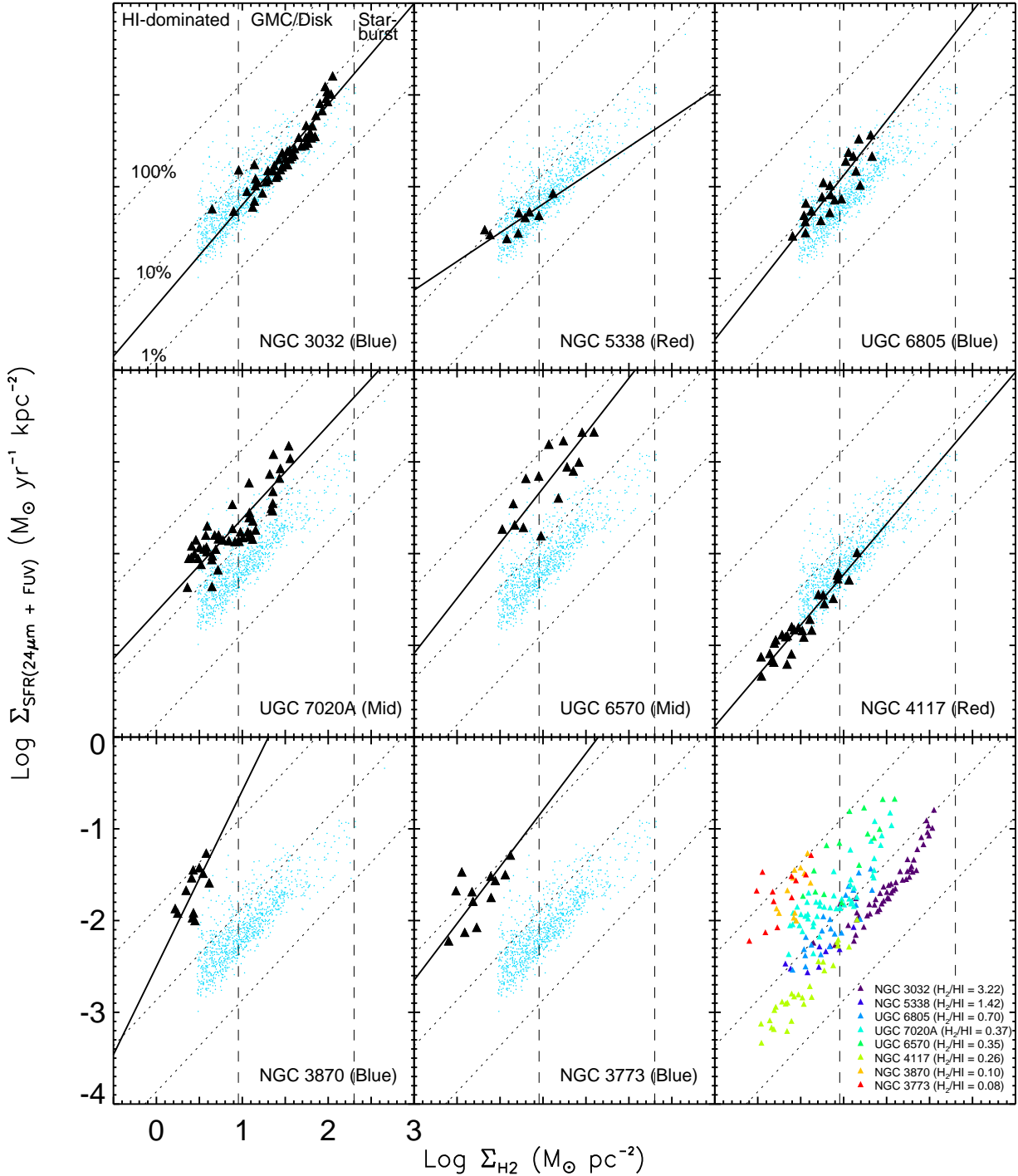}
\caption{${\rm \Sigma_{SFR}}$ vs. ${\rm \Sigma_{H_2}}$ at 750\,pc
  resolution. Points for the seven spirals from \citet{bigiel08} are
  plotted in light blue. Black triangles show our eight E/S0s resolved
  in CO(1--0) with CARMA. Vertical dashed lines demarcate the three
  different regimes of star formation discussed in \citet{bigiel08},
  and dotted lines mark constant MSFE, corresponding (from top to
  bottom) to the depletion of 100\%, 10\%, and 1\% of the molecular
  gas (including helium) within $10^8$\,yr, or equivalently to
  molecular gas depletion timescales of 0.1, 1, and 10 Gyr. Solid
  black lines represent OLS bisector fits. The color sequence
  (blue/mid/red) is noted at the bottom of each panel. The last panel
  combines the points for all eight galaxies, color-coded by $M_{\rm
    H_{2}}$/$M_{\rm HI}$, illustrating variations in MSFE with
  power-law index $N$ and $M_{\rm H_{2}}$/$M_{\rm HI}$ (see \S
  \ref{section.pixels} and \S
  \ref{section.discussion}). \label{fig.ks_pixels}}.
\end{figure}

Figure~\ref{fig.ks_pixels} plots the pixel-to-pixel relationship
between $\Sigma_{\rm SFR}$ and $\Sigma_{\rm H_{2}}$ for the eight of
our 19 galaxies resolved on 750\,pc scales. We also include the
750\,pc resolution data for the seven spirals observed by
\citet{bigiel08} as light blue dots. The vertical dashed lines
demarcate the three different star formation regimes discussed by
\citet{bigiel08}: HI-dominated, giant molecular cloud (GMC)/disk, and
starburst. Figure~\ref{fig.ks_pixels} shows that all but two of the
eight galaxies have some regions that fall within the GMC/disk regime,
with two blue- and one mid-sequence E/S0s (NGC~3032, UGC~6570, and
UGC~7020A) having the majority of their points in this regime.

We fit the ${\rm \Sigma_{H_2}}$--${\rm \Sigma_{SFR}}$ relationship
with a power-law of the form

\begin{equation}
  \Sigma_{\rm SFR} = a\left(\frac{\Sigma_{\rm H_2}}{b\,M_{\odot}\,{\rm
        pc^{-2}}}\right)^N
\end{equation}

\noindent in log-log space using the ordinary least-squares (OLS)
bisector method (solid line in Figure~\ref{fig.ks_pixels}) and list
the fit parameters in Table \ref{table.ksfits}, using coefficient
$A\,=\,{\rm log}_{10}\,(a/{\rm M_\odot\,kpc^{-2}\,yr^{-1}})$. Note
that we set the intercept of our fit at log$_{10}$($b$), where $b$ is
the median ${\rm \Sigma_{H_2}}$ for each galaxy, to lessen the effect
of the covariance between $N$ and $A$. The power-law index \textit{N}
ranges from 0.62 to 1.92, with a median of $\sim\,1.2$. It is evident
that the majority of E/S0s in Figure~\ref{fig.ks_pixels} exhibit a
power-law relation between ${\rm \Sigma_{H_2}}$ and ${\rm
\Sigma_{SFR}}$, all the way down to the HI-dominated regime.

Figure~\ref{fig.ks_pixels} also plots (dotted) lines of constant MSFE
with $N = 1$, defined as the inverse of the molecular gas depletion
time $M_{\rm H_2+He}$/SFR, and illustrates variations in $N$ and
$M_{\rm H_2}$/$M_{\rm HI}$. For blue- and mid-sequence E/S0s, we find
that as MSFE increases, $N$ seems to steepen and $M_{\rm H_2}$/$M_{\rm
  HI}$ seems to decrease. We find a wider range of MSFEs (4\%--70\%)
compared to the 3\%--8\% found by \citet{bigiel08}. We discuss whether
MSFE truly measures molecular-gas star formation \textit{efficiency}
in \S \ref{section.discussion}.

\section{The Global Star Formation Relation}\label{section.global}

In Figure \ref{fig.ks_global}, we show the 1--8 kpc aperture-averaged
relationship between $\Sigma_{\rm H_2}$ and $\Sigma_{\rm SFR}$ for the
eight E/S0s of Figure \ref{fig.ks_pixels} plus 11 more blue-sequence
E/S0s with CARMA detections and/or IRAM observations, as well as for
normal spiral-disk and starburst galaxies \citep{kennicutt98} and
SAURON E/S0s \citep{shapiro10}.  With the exception of Kennicutt's
normal disk points, all the other points in Figure~\ref{fig.ks_global}
show the surface densities within regions of star formation or
molecular gas. The starbursts were averaged over ``the radius of the
starburst region'' determined from CO/infrared imaging
\citep{kennicutt98}, and the SAURON E/S0s were averaged over the
extent of the star-forming region defined by 8\,$\mu$m PAH emission
\citep{shapiro10}. For our 19 E/S0s, we average $M_{\rm H_2}$ and SFR
over an area with radius ($R_{\rm ap}$) twice the scale length of the
24\,$\mu$m emission, where the flux drops by e$^{-2}$ from the
peak. This area encompasses most of the flux in CO, 24\,$\mu$m, and
FUV, and corresponds well to the visual impression of the extent of CO
emission for most galaxies. Note that the dots from \citet{bigiel08}
plotted in the background for reference are \textit{local} (750\,pc)
measures of ${\rm \Sigma_{H_2}}$ and ${\rm \Sigma_{SFR}}$ as in
Figure~\ref{fig.ks_pixels}.

In contrast, the \citet{kennicutt98} spiral disk points are averaged
over $R_{25}$, which may dilute the values of ${\rm \Sigma_{H_2}}$ and
${\rm \Sigma_{SFR}}$.  We infer this from radial profiles of normal
spirals in \citet{bigiel08}, which indicate that H$_2$ typically
extends out to only $\sim0.6\,R_{25}$. Thus the surface densities
calculated by \citet{kennicutt98} for spiral disks should for
consistency move up along lines of constant MSFE by $\sim$\,0.62 dex
(grey arrow in Figure~\ref{fig.ks_global}).

Comparison of these data sets reveals that the aperture-averaged ${\rm
  \Sigma_{H_2}}$ and ${\rm \Sigma_{SFR}}$ for our 19 E/S0s overlap the
range spanned by spiral disks, with six of the CARMA-detected E/S0s
having sufficiently high surface densities to occupy the same space
where \citet{kennicutt98} finds the \textit{centers} of spirals lie
--- between normal disks and starburst galaxies (not shown in
Figure~\ref{fig.ks_global}, as \citeauthor{kennicutt98} tabulates only
combined HI+H$_2$ data for galaxy centers). Our E/S0s also appear to
span the same range as the typically more massive SAURON E/S0s.

Figure~\ref{fig.ks_global} also shows a similarly large spread in
aperture-averaged MSFEs as seen for local MSFEs in \S
\ref{section.pixels}, with over half of the CARMA-detected and all of
the 30$\,$m-observed E/S0s (all blue- or mid-sequence) offset towards
apparently higher MSFEs ($>$10\%) compared to the typical spirals from
\citet{bigiel08}. Equivalently, the molecular gas depletion times for
our E/S0s range from 2.3 down to 0.1\,Gyr, with a median of 0.5\,Gyr
--- lower than that of the \citeauthor{bigiel08}
spirals. Additionally, the offset towards apparently higher MSFEs is
seen in both the CARMA and IRAM 30\,m galaxies in
Figure~\ref{fig.ks_global}, so it is not specific to interferometric
data.

\section{Discussion}\label{section.discussion}

We have shown above that the relationship between molecular gas and
star formation in low-mass E/S0s ($M_*\,\la4\times10^{10}\,M_{\odot}$)
resolved at 750\,pc is similar to that for spirals, with a roughly
linear correlation between ${\rm \Sigma_{H_2}}$ and ${\rm
  \Sigma_{SFR}}$ all the way down to the HI-dominated regime. This
suggests that star formation occurs in H$_2$ and not HI, similar to
what \citet{bigiel08} find for spirals. One intriguing difference is
the apparently elevated MSFEs of our E/S0s compared to the MSFEs of
the Bigiel spirals. \textit{Star formation efficiency}, however, may
be a misnomer in some cases, as other factors may contribute to the
observed offsets.

One possible cause for apparently elevated MSFEs is that the CO may
not trace all of the H$_2$ in these galaxies, as many are low-mass
systems where $X_{\rm CO}$ may be variable (e.g.,
\citealt{maloney88,pak98,pelupessy09}). \citet{kennicutt98} finds a
similar scatter in MSFEs for low-luminosity ($L_B < 10^{10}\,
{L_{\odot}}$) disk galaxies, which he attributes to variation in
$X_{\rm CO}$, possibly due to low metallicities. Recent
\textit{Herschel} results support this, finding evidence for excess
cold dust that is not well-traced by CO in low-mass galaxies
\citep{ohalloran10,kramer10}. However, metallicity measurements
(available for 5/6 CARMA-detected and 3/7 IRAM-observed galaxies
offset towards higher MSFEs) indicate that these galaxies are well
within the range ($\gtrsim\,$1/4\,$Z_{\odot}$) where much work
suggests that $X_{\rm CO}$ is similar to that assumed here
\citep[e.g.,][]{rosolowsky03,leroy06,wolfire10}.

A second explanation is that some of these galaxies are advanced,
H$_2$-depleted starbursts, where the delay between H$_2$ exhaustion
and fading of star-formation tracers associated with young, massive
stars gives the \textit{appearance} of elevated MSFEs. Thus the
possible correlation between elevated MSFE, lower $M_{\rm
H_2}$/$M_{\rm HI}$, and steeper (higher $N$) slopes in the
Schmidt-Kennicutt relation found in \S \ref{section.pixels} may
reflect the depletion of H$_2$ in the later stages of star formation,
in good agreement with simulations that predict steeper
Schmidt-Kennicutt relations and higher MSFEs as the molecular gas
fraction decreases \citep{robertson08}. This is consistent with
studies of NGC~1569, a post-starburst dwarf irregular (e.g.,
\citealt{angeretti05}) that also appears to have an elevated MSFE
\citep{leroy06}.

\begin{figure}
\hspace{-1cm}
\includegraphics[scale=1,angle=0]{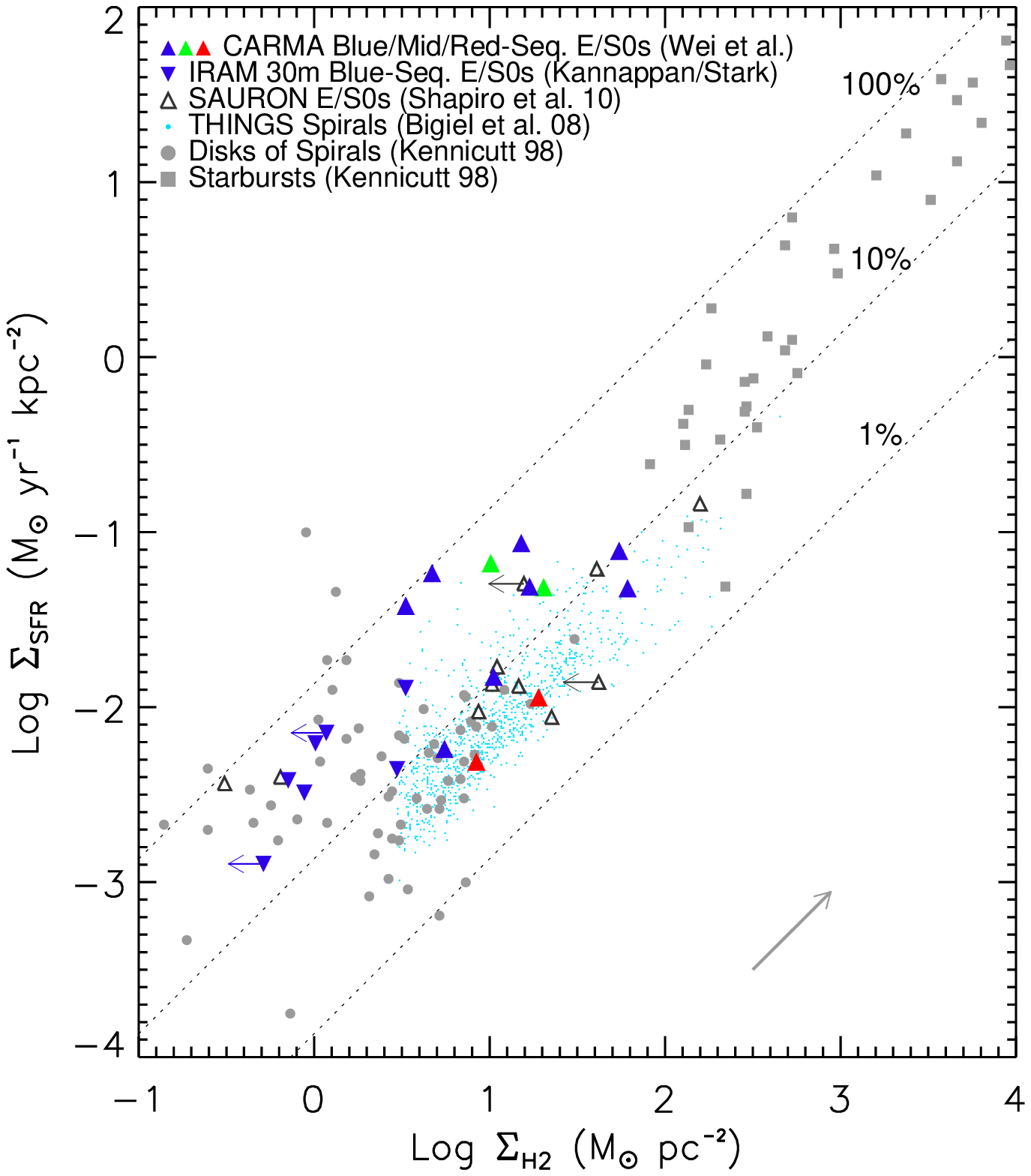}
\vspace{-1cm}
\caption{Aperture-averaged ${\rm \Sigma_{SFR}}$ vs. ${\rm
    \Sigma_{H_2}}$ for normal disk and nuclear starburst galaxies from
  \citet{kennicutt98}, SAURON E/S0s from \citet{shapiro10}, and E/S0s
  from this Letter. The 750\,pc resolution points from
  \citet{bigiel08} are also plotted in the background for
  comparison. Dotted lines mark the same lines of constant MSFE as in
  Figure~\ref{fig.ks_pixels}. Grey arrow in the lower right shows
  typical shift of spiral disk points from \citet{kennicutt98} if
  averaged over $0.6\,R_{25}$ instead of
  $R_{25}$. \label{fig.ks_global}}
\end{figure}

A third possibility is that the observed offsets may reflect truly
enhanced MSFEs. \citet{kannappan09conf}, updating \citet{kannappan04},
find that the fractional gas content of galaxies abruptly rises below
a gas-richness threshold mass of $M_*\,\sim\,3$--$5 \times
10^9\,M_{\odot}$, roughly corresponding to internal velocities of
$\sim$\,120 km\,s$^{-1}$ (see also \citetalias{kgb}). This is the same
velocity threshold below which the physics of star formation may
change due to possibly increased gas accretion, outflow, and metal
loss from shallower potentials
\citep{dalcanton04,garnett02,dalcanton07}. While the \textit{total}
SFEs of dwarf galaxies are low compared to those of normal
star-forming spirals
\citep[e.g.,][]{hunter04,dalcanton04,dalcanton07,robertson08}, recent
simulations suggest that gas-rich and/or lower metallicity galaxies
deviate from the Schmidt-Kennicutt relationship towards higher
\textit{molecular} SFEs \citep{pelupessy09}. These predictions are
supported by observations of local dwarfs, which find high MSFEs in
IC~10 and M33 compared to nearby spirals
\citep{leroy06,gardan07}. This phenomenon could contribute to the
scatter in MSFEs observed by \citet{kennicutt98}, as
$L_B\,\sim\,10^{10}\,{L_{\odot}}$ roughly corresponds to the
gas-richness threshold mass. Similarly, the two SAURON E/S0s in
Figure~\ref{fig.ks_global} with the highest MSFEs are low-luminosity
systems with $L_B < 10^{10}\, {L_{\odot}}$.

Unlike previous studies of star formation in blue E/S0s, our sample
focuses on galaxies below the gas-richness threshold mass (13 of our
19 E/S0s). At higher stellar masses, AGN and strong starbursts are
observed to dominate the blue E/S0 population
\citep[e.g.,][]{schawinski09,lee10}, which is not inconsistent with
the nature of our six higher-mass E/S0s (three known/candidate AGNs
and two likely starbursts\footnote{UGC~12265N \& UGC~6003 do not
appear in the starburst regime of Figure~\ref{fig.ks_global} due to
their large distances and the resulting poor resolution of the central
region, but their nuclear EW(H$\alpha$) emission measurements (86 and
76\,\AA, respectively) and very blue-centered color gradients suggest
recent/ongoing central starbursts.}).

Our results suggest that (possibly milder) bursts likely play a key
role for lower-mass E/S0s as well. The variation of MSFE in this
scenario has implications for our understanding (and theoretical
simulations) of low-mass galaxy evolution. The dynamical timescales
for gas inflow typical for our sample (0.06--0.4 Gyr; \citealt{wei10})
are short compared to the molecular gas consumption times we find here
(0.1--2.3 Gyr), which suggests that refueling of H$_2$ from the HI
reservoir to sustain star formation is limited only by the frequency
of minor mergers/interactions that trigger gas inflow. If apparently
elevated MSFEs reflect advanced, H$_2$-depleted bursts, the fact that
15 of our 19 E/S0s have MSFE $>$10\% suggests that low-mass E/S0s may
experience frequent small starbursts (with the caveat that we have
sampled the most strongly star-forming examples by favoring
blue-sequence E/S0s detected in CO). Therefore the TSFEs of lower-mass
galaxies, when time-averaged over many bursts, may be elevated ---
consistent with the \citetalias{kgb} finding that the concentration
indices in blue-sequence E/S0s are similar to those identified by
\citet{kauffmann06} as optimal for peak time-averaged TSFE. Our
results support the picture of \citetalias{kgb} and \citet{wei10} that
many blue-sequence E/S0s are in a ``sweet spot'' with abundant gas and
bursty star formation enabling efficient disk building.

%%%%%%%%%%%%%%%%%%%%%%%%%%%%%%%%%%%% ACKNOW. %%%%%%%%%%%%%%%%%%%%%%%%%%%%%%%%%%%%

\acknowledgements 

We thank the referee for helpful comments, A. Leroy for sharing the
THINGS data and for useful discussions, and S. Jogee for her role in
acquiring the \textit{Spitzer} data. We are grateful to A. Bolatto,
J. Gallimore, M. Lacy, A. Moffett, M. Thornley, and S. Veilleux for
insightful conversations. This work is based in part on observations
made with the \textit{Spitzer Space Telescope}, which is operated by
JPL, Caltech under a contract with NASA. Support for this work was
provided by NASA through an award issued by JPL/Caltech. This work
uses observations made with the NASA Galaxy Evolution
Explorer. \textit{GALEX} is operated for NASA by Caltech under NASA
contract NAS5-98034. We acknowledge support from the \textit{GALEX} GI
grant NNX07AT33G. CARMA development and operations are supported by
NSF under a cooperative agreement, and by the CARMA partner
universities.

%%%%%%%%%%%%%%%%%%%%%%%%%%%%%%%%%%%% BIB %%%%%%%%%%%%%%%%%%%%%%%%%%%%%%%%%%%%

%%%%%%%%%%%%%%%%%%%%%%%%%%%%%%%%%%%% FIGURES %%%%%%%%%%%%%%%%%%%%%%%%%%%%%%%%%%%%

%%%%%%%%%%%%%%%%%%%%%%%%%%% Figure 1 %%%%%%%%%%%%%%%%%%%%%%%%%%%%%%%%%                   
%%%%%%%%%%%%%%%%%%%%%%%%%%%%%%%%%%%%%%%%%%%%%%%%%%%%%%%%%%%%%%%%%%%%%%                   

%%%%%%%%%%%%%%%%%%%%%%%%%%% Figure 2 %%%%%%%%%%%%%%%%%%%%%%%%%%%%%%%%%                   
%%%%%%%%%%%%%%%%%%%%%%%%%%%%%%%%%%%%%%%%%%%%%%%%%%%%%%%%%%%%%%%%%%%%%%                   

%%%%%%%%%%%%%%%%%%%%%%%%%%% Figure 3 %%%%%%%%%%%%%%%%%%%%%%%%%%%%%%%%%                   
%%%%%%%%%%%%%%%%%%%%%%%%%%%%%%%%%%%%%%%%%%%%%%%%%%%%%%%%%%%%%%%%%%%%%%                   

%%%%%%%%%%%%%%%%%%%%%%%%%%%%%%%%%%%% TABLES %%%%%%%%%%%%%%%%%%%%%%%%%%%%%%%%%%%%

%%%%%%%%%%%%%%%%%%%%%%%%%%%% Table 1 %%%%%%%%%%%%%%%%%%%%%%%%%%%%%%%%%                   

\begin{deluxetable}{llccccccccccc}
  \tabletypesize{\scriptsize}
  \rotate
  \tablewidth{0pt}
  \tablecaption{Aperture-Averaged Properties for CARMA \& IRAM 30\,m E/S0s\label{table.global}}
  
  \tablehead{ \colhead{Galaxy} & \colhead{Seq.} & \colhead{$D_{\rm
    maj}$} & \colhead{Dist.} &
    \colhead{$M_*$} & \colhead{$M_{\rm HI}$} &
    \colhead{$M_{\rm H_2,C}$} & \colhead{$M_{\rm H_2,S}$}
    &\colhead{$R_{\rm ap}$} & \colhead{inclin.}& \colhead{$\Sigma_{\rm
        H_{2}}$} &
    \colhead{$\Sigma_{\rm SFR}$} & \colhead{$\tau_{\rm dep}$} \\
    \colhead{} & \colhead{} & \colhead{($\arcmin$)} & \colhead{(Mpc)} &
    \colhead{(log\,$M_{\odot}$)} & \colhead{(log\,$M_{\odot}$)} &
    \colhead{(log\,$M_{\odot}$)} & \colhead{(log\,$M_{\odot}$)} &
    \colhead{($\arcsec$)} & \colhead{($^\circ$)} &
    \colhead{($M_{\odot}\,$pc$^{-2}$)} &
    \colhead{($M_{\odot}\,$yr$^{-1}\,$kpc$^{-2}$)} & \colhead{(Gyr)}}

  \startdata
  NGC 3011   & B & 0.8 & 25.7 &  9.4 &  8.3 &      &  $<$7.0 & 12.1 & 27 & $<$1.2       & 0.007$\pm$0.001 & $<$0.22 \\      
  NGC 3032   & B & 1.3 & 25.2 &  9.6 &  8.3 &  8.6 &  8.6\tablenotemark{a} & 10.6 & 26 & 61.1$\pm$2.0 & 0.048$\pm$0.002 & 1.74$\pm$0.10 \\
  NGC 3073   & B & 1.1 & 21.1 &  9.1 &  8.5 &      &  7.0\tablenotemark{b} & 10.6 & 31 &  3.0$\pm$0.6 & 0.004$\pm$0.001 & 0.91$\pm$0.27 \\
  UGC 6003   & B & 0.6 & 84.2 & 10.1 &  9.4 &  8.6 &  8.7 &  7.5 & 19 & 15.1$\pm$1.4 & 0.087$\pm$0.003 & 0.24$\pm$0.02 \\
  IC 692     & B & 0.8 & 21.4 &  8.9 &  8.4 &      &  6.8 & 12.1 & 42 &  1.0$\pm$0.3 & 0.006$\pm$0.001 & 0.22$\pm$0.08 \\
  UGC 6570\tablenotemark{*}   & M & 1.2 & 28.6 &  9.6 &  8.4 &  8.0 & 8.0 &  7.5 & 62 & 10.1$\pm$0.8 & 0.066$\pm$0.001 & 0.21$\pm$0.02 \\
  UGC 6637\tablenotemark{*}   & B & 0.9 & 31.5 &  9.2 &  8.6 &      & 7.3 & 12.1 & 66 &  0.7$\pm$0.2 & 0.004$\pm$0.001 & 0.25$\pm$0.07 \\
  NGC 3773   & B & 1.1 & 10.5 &  8.6 &  7.9 &  6.6 &      & 10.6 & 44 &  4.7$\pm$0.6 & 0.058$\pm$0.004 & 0.11$\pm$0.02 \\
  NGC 3870   & B & 1.0 & 14.5 &  8.8 &  8.4 &  6.9 &  7.2\tablenotemark{c} & 12.1 & 41 &  3.3$\pm$0.5 & 0.038$\pm$0.003 & 0.12$\pm$0.02 \\
  UGC 6805   & B & 0.7 & 20.3 &  8.9 &  7.6 &  7.6 &  7.4 &  9.0 & 45 & 10.5$\pm$0.9 & 0.015$\pm$0.001 & 0.96$\pm$0.12 \\
  UGC 7020A  & M & 1.1 & 26.7 &  9.3 &  8.6 &  8.3 &  8.2 &  7.5 & 62 & 20.3$\pm$1.2 & 0.048$\pm$0.002 & 0.57$\pm$0.04 \\
  NGC 4117   & R & 1.8 & 19.0 &  9.7 &  8.3 &  7.7 &      &  7.5 & 72 &  8.4$\pm$0.6 & 0.005$\pm$0.000 & 2.35$\pm$0.25 \\
  NGC 5173   & B & 1.2 & 41.2 & 10.3 &  9.3 &  8.2 &      & 10.6 & 39 &  5.5$\pm$0.6 & 0.006$\pm$0.001 & 1.30$\pm$0.23 \\
  NGC 5338   & R & 1.9 & 10.3 &  8.9 &  7.3 &  7.3 &  7.4\tablenotemark{d} &  7.5 & 57 & 19.0$\pm$1.2 & 0.011$\pm$0.001 & 2.27$\pm$0.26 \\
  UGC 9562   & B & 0.9 & 25.2 &  8.9 &  9.3 &      &  7.0 & 12.1 & 68 &  0.9$\pm$0.3 & 0.003$\pm$0.001 & 0.37$\pm$0.15 \\
  IC 1141    & B & 0.7 & 68.0 & 10.4 &  9.3 &  9.2 &      &  7.5 & 26 & 54.7$\pm$2.7 & 0.078$\pm$0.002 & 0.95$\pm$0.06 \\
  NGC 7077\tablenotemark{*}   & B & 0.9 & 18.9 &  8.8 &  8.2 &      & 7.1 & 10.6 & 40 &  3.3$\pm$0.5 & 0.013$\pm$0.002 & 0.35$\pm$0.07 \\
  NGC 7360   & B & 1.2 & 67.9 & 10.5 &  9.6 &      &  $<$7.8 & 12.1 & 67 & $<$0.5       & 0.001$\pm$0.000 & $<$0.55       \\
  UGC 12265N & B & 0.6 & 82.8 & 10.1 &  9.4 &  8.9 &  8.9 &  7.5 & 42 & 16.9$\pm$1.3 & 0.049$\pm$0.001 & 0.47$\pm$0.04
  \enddata

  \tablenotetext{*}{Inclination assumed in calculating ${\rm
      \Sigma_{H_2}}$ and ${\rm \Sigma_{SFR}}$ is uncertain. }

  \tablecomments{Optical major axis ($D_{\rm maj}$), $M_*$, and
    distance data are from \citetalias{kgb}. HI data are from
    \citet{wei10} and references therein.  $M_{\rm H_2,C}$ is
    estimated from CARMA CO(1--0) data, and $M_{\rm H_2,S}$ from
    single-dish data (central pointings only) from Kannappan et al.,
    in prep. and Stark et al., in prep., except as marked: (a)
    \citet{thronson89}, (b) \citet{sage07}, (c) \citet{welch03}, (d)
    \citet{leroy05}. $\tau_{\rm dep} \equiv M_{\rm H_2+He}$/SFR.}

\end{deluxetable}
%%%%%%%%%%%%%%%%%%%%%%%%%%%%%%%%%%%%%%%%%%%%%%%%%%%%%%%%%%%%%%%%%%%%%%                   

%%%%%%%%%%%%%%%%%%%%%%%%%%%% Table 2 %%%%%%%%%%%%%%%%%%%%%%%%%%%%%%%%%                   

\begin{deluxetable}{lccccc}
  \tablewidth{0pt}
  \tablecaption{Star Formation Relation Fit Parameters\label{table.ksfits}}

  \tablehead{ \colhead{Galaxy} & \colhead{\textit{b}} &
    \colhead{Coeff. \textit{A}} & \colhead{Index \textit{N}} &
    \colhead{RMS} \\ \colhead{} & \colhead{($M_{\odot}\,{\rm pc^2}$)} &
    \colhead{} & \colhead{} & \colhead{(dex)} }

  \startdata
  \textbf{Blue Seq.:} & & & &  \\
  NGC~3032  &  33.7  & -1.62$\pm$0.02  & 1.10$\pm$0.08  & 0.12\\
  NGC~3773  &   1.7  & -1.75$\pm$0.07  & 1.24$\pm$0.16  & 0.20\\
  NGC~3870  &   2.7  & -1.67$\pm$0.06  & 1.92$\pm$0.35  & 0.23\\
  UGC~6805  &   7.1  & -2.06$\pm$0.03  & 1.19$\pm$0.10  & 0.18\\
  \hline
  \textbf{Mid Seq.:} & & & &  \\
  UGC~6570  &  11.7  & -1.21$\pm$0.06  & 1.20$\pm$0.14  & 0.23\\
  UGC~7020A &   7.7  & -1.73$\pm$0.03  & 1.02$\pm$0.08  & 0.21\\
  \hline
  \textbf{Red Seq.:} & & & &  \\
  NGC~4117  &   3.0  & -2.81$\pm$0.02  & 1.10$\pm$0.06  & 0.13\\
  NGC~5338  &   5.2  & -2.37$\pm$0.03  & 0.62$\pm$0.11  & 0.09\\
  \hline
  Median    &   7.1  & -1.73           & 1.19           &     
  \enddata

\end{deluxetable}
%%%%%%%%%%%%%%%%%%%%%%%%%%%%%%%%%%%%%%%%%%%%%%%%%%%%%%%%%%%%%%%%%%%%%%                   

\end{document}